\documentclass[preprint,preprintnumbers,amsmath,amssymb]{revtex4}

\usepackage{graphicx}
\usepackage{dcolumn}
\usepackage{bm}
\begin{document}
\title{Transport in a spin polarized Luttinger liquid in the presence of
a single impurity}
\author{Hsuan-Yeh Chang and D. Schmeltzer}
\affiliation{Physics Department, City College of the City University of New York \\
New York, New York 10031}

\date{\today}
\begin{abstract}
The conductance for a single impurity in a one dimensional gas
with electron-electron interaction and Zeeman splitting
$v_F^\uparrow - v_F^\downarrow = \Delta \neq 0$ is considered.  In order to solve the model Hamiltonian for the entire spectrum, we need to transform our model into the Lagrangian formalism.  We find that for an extented Hubbard model, i.e. in the presence of the $\rho_\uparrow^R \rho_\uparrow^L + \rho_\downarrow^R \rho_\downarrow^L$ term, a Luttinger liquid for spin up and spin down is obtained.  As a result, the conductances for spin up $G_\uparrow$ and spin down $G_\downarrow$ are not equal.  The anisotropy in the conductances $G_\uparrow \neq G_\downarrow$ gives rise to a spin polarized wire.
\end{abstract}

\maketitle
\section{Introduction}
Recently, a new kind of electronics named ``spintronics'' based on spin polarized electrons have created a wave of excitement [1].  The idea to build a spin diode or a spin transistor is extremely appealing.  In order to make progress, one has to understand ``spin ballistic transport''.

Here, we will investigate the possibility for having polarized
electrons in a quantum wire.  We compute the conductance of a
non-magnetic impurity in the presence of electron-electron
interaction with two different Fermi velocities $v_\uparrow \neq
v_\downarrow$.  Our findings are the followings:

(a) In the presence of a Hubbard type interaction, the different
Fermi velocities $v_\uparrow \neq v_\downarrow$ do not give rise
to a polarized gas.  In particular, the conductance for the
non-magnetic impurity is given by the Luttinger theory in terms of
spin and charge separated parameters.  If $\hat{\lambda}$ is the
strength of the ``$2 k_F$'' impurity potential and ``$K_c < 1$'',
``$K_s \approx 1$'' are the Luttinger parameters, the conductance
$G_\sigma$ is given by [2], $G_\uparrow = G_\downarrow =
\frac{e^2}{h} \left(1 - const. \hat{\lambda}^2 ( \frac{T}{T_F}
)^{2(K_c - 1 )} \right) {}_{\overrightarrow{T \rightarrow 0}} 0$.

(b) For special types of two body interactions characteristic to
extended Hubbard model of the form, $\rho_\uparrow^R(x)
\rho_\uparrow^L(x) + \rho_\downarrow^R(x) \rho_\downarrow^L(x)$,
we find that in the presence of two Fermi velocities $v_\uparrow
\neq v_\downarrow$ one obtains a new type of Luttinger liquid, a
non-spin-charge separated liquid characterized by two exponents
$K_\uparrow \neq K_\downarrow$ and not $K_c \neq K_s$.  As a
result we find that the conductance $G_\sigma$ obeys, $G_\uparrow
\neq G_\downarrow$ with $G_\sigma = \frac{e^2}{h} \left(1 - const.
\hat{\lambda}^2 (\frac{T}{T_F})^{2(K_\sigma-1)} \right) = G_\sigma^{(0)} + G_\sigma^{(2)}$. For
strong anisotropy, $1 \approx K_\uparrow \gg K_\downarrow$ we
obtain at finite temperatures $G_\downarrow \approx 0$ and
$G_\uparrow \approx \frac{e^2}{h}$, therefore a polarized wire is
obtained.

These results are relevant to ``spintronics'' [1], since it shows
that the interplay between interaction, impurities and different
Fermi velocities can give rise to a polarized current, $P =
\frac{G_\uparrow - G_\downarrow}{G_\uparrow + G_\downarrow}$.  The plan of this paper is the following: chapter II is devoted to the microscopic model, chapter III is devoted to the numerical calculation of $K_\uparrow$, $K_\downarrow$ and the conclusions.

\section{The Model}
Using the 1-d Fermion representation $\psi_\sigma(x) = e^{i
K_{F,\sigma} x} R_\sigma(x) + e^{-iK_{F,\sigma}x}L_\sigma(x)$,
$\sigma = \uparrow, \downarrow$ where $R_\sigma(x)$ and
$L_\sigma(x)$ represent the right and left mover [1, 2-6].  We
consider a situation that due to the Zeeman interaction the Fermi
gas is described by $v_F^{\uparrow} = v_F + \frac{\Delta}{2}$,
$v_F^{\downarrow} = v_F - \frac{\Delta}{2}$, $k_F^{\sigma}/m =
v_F^{\sigma}$, $\sigma = \uparrow, \downarrow$.  The parameter
$\Delta$ represents the Zeeman energy.  We consider the following one dimensional electronic model: $H = H_0 + H_I + H_{Imp}$, where $H_0 = - t \sum_x \sum_{\sigma = \uparrow, \downarrow} [ \psi_\sigma^\dag(x) \psi_\sigma(x+a) + h.c. ]$, $H_I = U \sum_x n_\uparrow(x) n_\downarrow (x) + V \sum_x n(x) n(x+a)$.  Here, ``$U$'' is the on-site (Hubbard) interaction and $n_\sigma(x) = \psi_\sigma^\dag(x) \psi_\sigma(x)$, while ``$V$'' is the nearest neighbor (extended Hubbard) interaction and $n(x) = \sum_\sigma n_\sigma(x)$.  For the $2 k_F$ impurity, $H_{imp} = \sum_x U_\sigma(x) \psi_\sigma^\dag(x) \psi_\sigma(x)$, where $U_\sigma(x)$ is the $2 k_F$ impurity potential.  As a result, the Hamiltonian for the 1-d interacting Fermi gas with the $2 k_F$ impurity takes the form $H = H_0 + H_I + H_{imp}^{(2 k_F)}$,
\begin{eqnarray}
H_0 &=& - \sum_{\sigma= \uparrow, \downarrow}\int dx
v_F^{(\sigma)}\left[ R_\sigma^\dag(x) i \partial_x R_\sigma(x) -
L_\sigma^\dag(x) i
\partial_x L_\sigma(x)\right] \\ \nonumber
H_I &=& \sum_{\sigma, \sigma' = \uparrow, \downarrow} \int dx
\left\{ g_1 R_\sigma^\dag(x) L_{\sigma'}^\dag(x) R_{\sigma'}(x)
L_\sigma(x) + g_2 R_\sigma^\dag(x) L_{\sigma'}^\dag(x)
L_{\sigma'}(x)
R_\sigma(x) \right. \\
&& \hspace{2cm} + \left. g_4 ( 1 - \delta_{\sigma, \sigma'})
\left[ R_\sigma^\dag(x) R_{\sigma'}^\dag(x) R_{\sigma'}(x)
R_\sigma(x) + \; (R \leftrightarrow L) \; \right] \; \right\} \\
H_{Imp}^{(2 k_F)} &=& \sum_{\sigma=\uparrow, \downarrow} \int dx
\, U_\sigma(x) \left[ e^{-i 2 k_{F,\sigma}}
R_\sigma^\dag(x)L_\sigma(x) + h.c. \right]
\end{eqnarray}
where $R_\sigma$ and $L_\sigma$ are the right and left mover,
respectively.  Here, the $g_1$ term represents the backward
scattering which transfers momentum from $(k_F, -k_F)$ to $(- k_F,
k_F)$.  The $g_2$ term represents the forward scattering: $(- k_F,
k_F) \rightarrow (-k_F, k_F)$. [3-5]

Using $R_\sigma(x) = \frac{1}{\sqrt{2 \pi a}} e^{i \sqrt{4
\pi}\theta_{R,\sigma}(x)}$ and $L_\sigma(x) = \frac{1}{\sqrt{2 \pi
a}} e^{-i\sqrt{4 \pi} \theta_{L,\sigma}(x)}$, we bosonize the
Hamiltonian in eqs 1-3.
\begin{eqnarray}
H_0 &=& \sum_{\sigma=\uparrow, \downarrow} \int dx \,
v_F^{(\sigma)} \left[ \left(\partial_x \theta_{R,
\sigma}(x)\right)^2 + \left(\partial_x \theta_{L,
\sigma}(x)\right)^2 \right] \\ \nonumber H_I &=& \int dx \left\{
\left(\frac{g_2^\parallel - g_1^\parallel}{2 \pi}\right)
\left[(\partial_x \theta_{R \uparrow}(x))(\partial_x\theta_{L
\uparrow}(x)) + (\partial_x \theta_{R \downarrow}(x))(\partial_x
\theta_{L \downarrow}(x)) \right] \right. \\ \nonumber &&
\hspace{1.5cm} + \left(\frac{g_2^\bot}{2 \pi}\right)
\left[(\partial_x \theta_{R \uparrow}(x))(\partial_x\theta_{L
\downarrow}(x)) + (\partial_x \theta_{R \downarrow}(x))(\partial_x
\theta_{L \uparrow}(x)) \right] \\ \nonumber && \hspace{1.5cm}+
\left( \frac{g_4}{2 \pi}\right) \left[ (\partial_x \theta_{R
\uparrow}(x))(\partial_x\theta_{R \downarrow}(x) ) + (\partial_x
\theta_{L \uparrow}(x))(\partial_x \theta_{L \downarrow}(x))
\right] \\ && \hspace{1cm}+ \left. \left(\frac{g_1^\bot}{2 (\pi
d)^2}\right) \cos{\left[ \sqrt{4 \pi}(\theta_{R \uparrow} +
\theta_{L \uparrow} - \theta_{R \downarrow} -
\theta_{L\downarrow}) + \Delta (\frac{k_F}{v_F}) x \right]}
\right\} \\
H_{Imp}^{(2 k_F)} &=& \sum_{\sigma=\uparrow, \downarrow} \int dx
\, \lambda_\sigma \delta(x) \cos\left(\sqrt{4 \pi}
(\theta_{R,\sigma}(x) + \theta_{L,\sigma}(x) ) \right) , \quad
\lambda_\uparrow = \lambda_\downarrow
\end{eqnarray}

The parameter $g_2^\parallel$, $g_2^\bot$, $g_1^\parallel$, $g_1^\bot$ and $g_4$ are related to the Hubbard interaction ``$U$'' and intersite interaction, $g_2^\parallel = g_2^\bot \equiv U + 2 V$, $g_1^\parallel = g_1^\bot = U + 2 V \cos(k_F a) \approx U - 2 V$, and $g_2^\parallel - g_1^\parallel = 4 V$.

For a Hubbard model $V=0$, we have $g_2^\parallel - g_1^\parallel = 0$.  The presence of the parameter $\Delta(k_F/v_F) x$, $k_F =
(k_F^\uparrow + k_F^\downarrow)/2$ gives rise to strong
oscillation which allows us to neglect the last term in eq. 5 at
large distances [3].  $v_F^\sigma$ is the velocity corresponding to
different spin components $\sigma$.  In eq 6, we have used
$U_\sigma(x) = \lambda_\sigma \delta(x)$ with  $\lambda_\uparrow =
\lambda_\downarrow$, which represent a non-magnetic impurity located at $x=0$.  By applying an external magnetic field to the system, we can get
two different velocities for different spins, i.e. $v_\uparrow =
v_F + \Delta/2$ and $ v_\downarrow = v_F - \Delta/2$, where
$\Delta$ is the Zeeman energy.  For a next nearest neighbor interaction $V \neq 0$, $g_2^\parallel - g_1^\parallel = 4 V \neq 0$.  We denote $g_2^\parallel - g_1^\parallel \equiv g_0$ for simplicity.  Notice that this $g_0$ term is the only parameter relevant to the polarization of the
magnetic wire.  All other terms are not sensitive to the
externally applied magnetic field.  It is equivalent to the
inclusion of the nearest-neighbor interaction.

Now we transform the right and left moving fields into spin-up and
spin-down fields: $\theta_{R \sigma} = \frac{1}{2}(\theta_\sigma +
\phi_\sigma)$, $\theta_{L \sigma} = \frac{1}{2}(\theta_\sigma -
\phi_\sigma)$. We can further transform the spin-up and spin-down
fields into spin and charge fields: $\theta_{c} =
\frac{1}{\sqrt{2}}(\theta_\uparrow + \theta_\downarrow)$,
$\theta_{s} = \frac{1}{\sqrt{2}}(\theta_\uparrow -
\theta_\downarrow)$, $\phi_{c} = \frac{1}{\sqrt{2}}(\phi_\uparrow
+ \phi_\downarrow)$, $\phi_{s} = \frac{1}{\sqrt{2}}(\phi_\uparrow
- \phi_\downarrow)$. Let $P_c(x)=\partial_x \phi_c(x)$, and
$P_s(x)=\partial_x \phi_s(x)$, after some algebra, our Hamiltonian
becomes $H(P_c, \theta_c; P_s, \theta_s)=\int dx \, h(P_c,
\theta_c; P_s, \theta_s) = H_c + H_s + H_{c/s}^Z + H_{Imp}$ where
\begin{eqnarray}
H_c &=& \int dx \frac{v_c}{2} \left[ K_c P_c^2(x) +
\frac{1}{K_c}(\partial_x \theta_c)^2 \right] \\
H_s &=& \int dx \frac{v_s}{2} \left[ K_s P_s^2(x) +
\frac{1}{K_s}(\partial_x \theta_s)^2 \right] \\
H_{c/s}^Z &=& \frac{\Delta}{2} \int dx \left[ P_c(x) P_s(x) +
(\partial_x \theta_c)(\partial_s \theta_s) \right]
\end{eqnarray}
and
\begin{equation} \nonumber
K_c = \sqrt{\frac{1 - \frac{1}{4\pi} (g_0 + g_2^\bot - g_4) }{1 +
\frac{1}{4 \pi}(g_0 + g_2^\bot + g_4)}}, \quad v_c =  \sqrt{ (1 -
\frac{1}{4\pi} (g_0 + g_2^\bot - g_4) ) ( 1 + \frac{1}{4 \pi}(g_0
+ g_2^\bot + g_4))} \end{equation}
\begin{equation} \nonumber
K_s = \sqrt{\frac{1 - \frac{1}{4\pi} (g_0 - g_2^\bot + g_4) }{1 +
\frac{1}{4 \pi}(g_0 - g_2^\bot - g_4)}}, \quad v_s =  \sqrt{ (1 -
\frac{1}{4\pi} (g_0 - g_2^\bot + g_4) ) ( 1 + \frac{1}{4 \pi}(g_0
- g_2^\bot - g_4))}
\end{equation}
We keep $H_{Imp}$ in its original form of equation (6) and perform
the Renormalization Group analysis [6].

In order to solve the Hamiltonian in equation (7),(8) and (9) for
the entire spectrum, we need to transform our model into the
Lagrangian formalism.  One can still solve this problem by diagonalizing the above Hamiltonian [7], but the solution is subjected to certain restrictions.  The Lagrangian density $\mathcal{L}$ is defined as
\begin{equation}
\mathcal{L}(\theta_c, \dot{\theta}_c, \theta_s, \dot{\theta}_s) =
P_c \dot{\theta}_c + P_s \dot{\theta}_s - h(P_c, \theta_c, P_s,
\theta_s)
\end{equation}
Since $\dot{\theta}_c = \frac{\delta}{\delta P_c} H$ and
$\dot{\theta}_s = \frac{\delta}{\delta P_s} H $, we have
$\dot{\theta}_c = v_c K_c P_c + \frac{\Delta}{2} P_s$ and
$\dot{\theta}_s = v_s K_s P_s + \frac{\Delta}{2} P_c$.  We can
write $\left(%
\begin{array}{c}
  \dot{\theta}_c \\
  \dot{\theta}_s \\
\end{array}%
\right) = \textbf{M}
\left(%
\begin{array}{c}
  P_c \\
  P_s \\
\end{array}%
\right)$, where $\textbf{M} = \left(%
\begin{array}{cc}
  v_c K_c & \frac{\Delta}{2} \\
  \frac{\Delta}{2} & v_s K_s \\
\end{array}%
\right)$. Therefore, $\left(%
\begin{array}{c}
  P_c \\
  P_s \\
\end{array}%
\right)  = \textbf{M}^{-1}
\left(%
\begin{array}{c}
  \dot{\theta}_c \\
  \dot{\theta}_s \\
\end{array}%
\right)
= \frac{1}{\det\textbf{M}} \left(%
\begin{array}{cc}
   v_s K_s & -\frac{\Delta}{2} \\
  -\frac{\Delta}{2} & v_c K_c \\
\end{array}%
\right)
\left(%
\begin{array}{c}
  \dot{\theta}_c \\
  \dot{\theta}_s \\
\end{array}%
\right)$.  Let $ z^{-1} = \det \textbf{M}  = (v_c K_c)(v_s K_s) -
(\Delta/2)^2$, we should require that $z^{-1} \neq 0$.  Notice that the $z^{-1} = 0$ case corresponds to the fact that the fields $\dot{\theta}_c$ and $\dot{\theta}_s$ (or $P_c$ and $P_s$) are linear dependent.

Provided the above discussion, after some algebra, we obtain the Lagrangian density as
\begin{displaymath}
\mathcal{L}(\theta_c, \dot{\theta}_c, \theta_s, \dot{\theta}_s) =
\frac{1}{2} (\dot{\theta}_c, \dot{\theta}_s) \cdot \textbf{M}^{-1} \cdot \left(%
\begin{array}{c}
  \dot{\theta}_c \\
  \dot{\theta}_s \\
\end{array}%
\right)
- \frac{1}{2} (\partial_x{\theta}_c, \partial_x{\theta}_s) \cdot \textbf{M} \cdot
\left(%
\begin{array}{c}
  \partial_x{\theta}_c \\
  \partial_x{\theta}_s \\
\end{array}%
\right).
\end{displaymath}
We can write the action in terms of Fourier components of the above Lagrangian density as
\begin{displaymath}
S = \int \frac{d \omega}{2 \pi} \frac{d q}{2 \pi}
\mathcal{L}(\omega, q)
\end{displaymath}
where
\begin{displaymath}
\mathcal{L}(\omega, q) =  \frac{1}{2} \, \vec{\theta}(\omega,q)
\cdot \left(\textbf{A}^{-1}\right) \cdot \vec{\theta}( -\omega,
-q)
\end{displaymath}
and
\begin{equation}
\textbf{A}^{-1}=\left(%
\begin{array}{cc}
  \alpha & \gamma \\
  \gamma & \beta \\
\end{array}%
\right)
\end{equation}
Here we denote $\vec{\theta} = (\theta_c, \theta_s)$, and the
elements of matrix $\textbf{A}^{-1}$ are given as
\begin{eqnarray} \nonumber
\alpha(\omega,q)&=& -(z v_s K_s ) \omega^2 + (\frac{v_c}{K_c})q^2
\\ \nonumber
\beta(\omega,q) &=& -(z v_c K_c ) \omega^2 + (\frac{v_s}{K_s})q^2 \\
\gamma(\omega,q)&=&\frac{\Delta}{2}(z \omega^2 + q^2)
\end{eqnarray}

In order to compute the conductance due to the impurity we add to
eqs 7-9 the impurity potential $H_{Imp}^{(2 k_F)} = \int dx \,
\lambda_\sigma \delta(x) \cos\left(\sqrt{2 \pi} (\theta_{c}(x) +
\sigma \theta_{s}(x) ) \right)$ with the condition of
$\lambda_\uparrow = \lambda_\downarrow$.  Due to the electron-electron interaction, the bare impurity strength get renormalized.  At length scale $a' = ba$ where $b = e^{\ell} > 1$,  we find that $\lambda_\sigma(\ell) = \hat{\lambda}_\sigma \Lambda \exp[(1- K_\sigma) \ell]$, i.e. when $K_\sigma < 1$, $\lambda_\sigma(\ell)$ increases.  On the contrary, $\lambda_\sigma$ decrease when $K_\sigma > 1$.  The scaling equation of
$\lambda_\sigma = \hat{\lambda}_\sigma \Lambda$ is
\begin{equation}
\frac{d \hat{\lambda}_\sigma}{d \ell} = \hat{\lambda}_\sigma ( 1 -
{K}_\sigma)
\end{equation}
and ${K}_\sigma$ is given by
\begin{equation}
d \ell {K}_\sigma \equiv \frac{1}{2}(\sqrt{2 \pi})^2 \left[
\langle \delta \theta_c^2(x) \rangle + \langle \delta
\theta_s^2(x) \rangle + 2 \sigma \langle \delta \theta_c(x) \delta
\theta_s(x) \rangle \right] \equiv 2 \pi \langle \delta
\theta_\sigma(x) \delta \theta_\sigma(x) \rangle
\end{equation}
Using the Lagrangian in eqs 11-12 we will compute the correlation
functions $\langle \delta \theta_\sigma(x) \delta \theta_\sigma(x)
\rangle$. These correlation functions will be expressed in terms
of the parameters $\alpha$,$\beta$,$\gamma$  (see eq. 12).

Denote that $i,j \equiv (c,s)$, we can transform the correlation
function back to the space-time coordinate. We have
\begin{eqnarray} \nonumber
\langle \theta_i(x,t) \theta_j(0) \rangle = \int \frac{d \omega}{2
\pi} \frac{d q}{2 \pi} e^{- i (q x - \omega t)} \langle
\theta_i(\omega,q) \theta_j(-\omega,-q) \rangle \\ = i\int \frac{d
\omega}{2 \pi} \frac{d q}{2 \pi} e^{- i (q x - \omega t)}
\textbf{A}_{ij}(\omega,q) \equiv i \textbf{A}_{ij}(x,t)
\end{eqnarray}
Since for any function $f(|q|)$, the integral in the momentum
shell gives $\int_{\Lambda/s \leq |q| \leq \Lambda} f(|q|) dq =
\int_{-\Lambda}^{-\Lambda/s} + \int_{\Lambda/s}^{\Lambda} = 2
f(\Lambda)$, we can obtain the correlation functions at the same
space point as
\begin{equation}
\langle \delta \theta_i(x,\tau) \delta \theta_j(x,0) \rangle =
\frac{i d \Lambda}{\pi} \int_{-\infty}^{\infty} \frac{d \omega}{2
\pi} e^{i \omega \tau}\textbf{A}_{ij}(\omega, \Lambda)
\end{equation}
Since $\delta \theta_\uparrow = \frac{1}{\sqrt{2}}(\delta \theta_c
+ \delta \theta_s)$ and $\delta \theta_\downarrow =
\frac{1}{\sqrt{2}}(\delta \theta_c - \delta \theta_s)$, we have
\begin{eqnarray}
\langle \delta \theta_\uparrow \delta \theta_\uparrow \rangle =
\frac{1}{2}\left[ \langle \delta \theta_c \delta \theta_c \rangle
+ \langle \delta \theta_s \delta \theta_s \rangle + \langle \delta
\theta_c \delta \theta_s \rangle \right] \\ \langle \delta
\theta_\downarrow \delta \theta_\downarrow \rangle =
\frac{1}{2}\left[ \langle \delta \theta_c \delta \theta_c \rangle
+ \langle \delta \theta_s \delta \theta_s \rangle - \langle \delta
\theta_c \delta \theta_s \rangle \right]
\end{eqnarray}
Or
\begin{eqnarray}
\langle \delta \theta_\uparrow \delta \theta_\uparrow \rangle =
\frac{i d \Lambda}{2 \pi} \int_{-\infty}^{\infty} \frac{d
\omega}{2 \pi} e^{i \omega \tau} f_{\uparrow\uparrow}(\omega,
\Lambda), \quad \textrm{where} \, f_{\uparrow\uparrow}(\omega,
\Lambda) \equiv \frac{\alpha + \beta
- 2 \gamma}{\alpha \beta - \gamma^2} \\
\langle \delta \theta_\downarrow \delta \theta_\downarrow \rangle
= \frac{i d \Lambda}{2 \pi} \int_{-\infty}^{\infty} \frac{d
\omega}{2 \pi} e^{i \omega \tau} f_{\downarrow\downarrow}(\omega,
\Lambda), \quad \textrm{where} \, f_{\downarrow\downarrow}(\omega,
\Lambda) \equiv \frac{\alpha + \beta + 2 \gamma}{\alpha \beta -
\gamma^2}
\end{eqnarray}

Now we need to carry out the definite integrals in equation (19)
and (20). These can be done by using the Cauchy Residue Theorem.
By following the requirement that $\alpha \beta - \gamma^2 = 0$,
we can obtain four different poles.  They are
$\omega_1 = \frac{\Lambda}{2}\sqrt{u_1 + u_2}$, $\omega_2 = -
\frac{\Lambda}{2}\sqrt{u_1 + u_2}$, $\omega_3 =
\frac{\Lambda}{2}\sqrt{u_1 - u_2}$, $\omega_4 = -
\frac{\Lambda}{2}\sqrt{u_1 - u_2}$, where $u_1 = 2 (v_c^2 + v_s^2)
+ \Delta^2$ and $u_2 = 2 [ ( v_c^2 - v_s^2)^2 + (K_c K_s v_c +
v_s)(K_s K_c v_s + v_c) \Delta^2 / (K_c K_s) ]^{1/2}$.  Notice that these poles are located on the real axis and that $\omega_2 = - \omega_1$ and $\omega_4 = - \omega_3$.  Due to the factor $e^{i \omega \tau}$ inside of both integrals, $\tau > 0$, we need to choose the contour around the upper hemisphere and shift $\omega_1$, $\omega_3$ by a small positive imaginary number and $\omega_2$, $\omega_4$ by a small negative imaginary number.
Thus, only the poles at $\omega_1$ and $\omega_3$ are included in the
chosen contour. We get
\begin{eqnarray} \nonumber
\langle \delta \theta_\uparrow \delta \theta_\uparrow \rangle = -
\frac{ d \Lambda}{2 \pi} \left[ Res\{ f_{\uparrow\uparrow}
(\omega_1)\}+ Res \{f_{\uparrow\uparrow}(\omega_3)\} \right] \\
\nonumber \langle \delta \theta_\downarrow \delta
\theta_\downarrow \rangle = - \frac{ d \Lambda}{2 \pi} \left[
Res\{ f_{\downarrow\downarrow} (\omega_1) \} + Res\{
f_{\downarrow\downarrow}(\omega_3)\} \right]
\end{eqnarray}

By carefully calculating the residues of $f_{\uparrow \uparrow}$
and $f_{\downarrow \downarrow}$, one can obtain that
\begin{displaymath}
\langle \delta \theta_\uparrow \delta \theta_\uparrow \rangle =
\frac{ d \Lambda}{2 \pi \Lambda} {K}_\uparrow = \frac{ d
\Lambda}{2 \pi \Lambda} \left[ \frac{a}{2 u_2}(\sqrt{u_1 + u_2} -
\sqrt{u_1 - u_2}) - \frac{2 b}{u_2}(\frac{1}{\sqrt{u_1 + u_2}} -
\frac{1}{\sqrt{u_1 - u_2}}) \right]
\end{displaymath}
where $a = (K_c v_c + K_s v_s + \Delta)$, $b=(\frac{v_s}{K_s} +
\frac{v_c}{K_c} - \Delta) / z$.  And
\begin{displaymath}
\langle\delta \theta_\downarrow \delta \theta_\downarrow \rangle =
\frac{ d \Lambda}{2 \pi \Lambda} {K}_\downarrow = \frac{ d
\Lambda}{2 \pi \Lambda} \left[ \frac{c}{2 u_2}(\sqrt{u_1 + u_2} -
\sqrt{u_1 - u_2}) - \frac{2 d}{u_2}(\frac{1}{\sqrt{u_1 + u_2}} -
\frac{1}{\sqrt{u_1 - u_2}}) \right]
\end{displaymath}
where $c = (K_c v_c + K_s v_s - \Delta)$, $ d = ( \frac{v_s}{K_s}
+ \frac{v_c}{K_c} + \Delta)/z$. We can now write the
${K}_\uparrow$ and ${K}_\downarrow$ as
\begin{eqnarray}
{K}_\uparrow &=& \frac{a}{2 u_2}(\sqrt{u_1 + u_2} - \sqrt{u_1 -
u_2}) - \frac{2 b}{u_2}(\frac{1}{\sqrt{u_1 + u_2}} -
\frac{1}{\sqrt{u_1 - u_2}}), \\ 
{K}_\downarrow &=& \frac{c}{2
u_2}(\sqrt{u_1 + u_2} - \sqrt{u_1 - u_2}) - \frac{2
d}{u_2}(\frac{1}{\sqrt{u_1 + u_2}} - \frac{1}{\sqrt{u_1 - u_2}}).
\end{eqnarray}
By increasing $\Delta$ (the magnetic field), we have found that
$K_\uparrow \neq K_\downarrow$ when $g_0 \neq 0$.  The conductances for spin up and spin down electrons due to a weak impurity (i.e. $\hat{\lambda}_\sigma$ small) is then given by
\begin{eqnarray} \nonumber
G_\uparrow = \frac{e^2}{h} \left( 1 - |\hat{\lambda}_\uparrow|^2
\left( \frac{T}{T_F} \right)^{2 ({K}_\uparrow-1)} \right),  \\
\nonumber G_\downarrow = \frac{e^2}{h} \left( 1 -
|\hat{\lambda}_\downarrow|^2 \left( \frac{T}{T_F}
\right)^{2({K}_\downarrow-1)} \right).
\end{eqnarray}
For a weak link, on the other hand, the conductances are given by [8]
\begin{eqnarray} \nonumber
G_\uparrow = \frac{e^2}{h} |\frac{1}{\hat{\lambda}_\uparrow}|^2
\left( \frac{T}{T_F} \right)^{2 (\frac{1}{{K}_\uparrow}-1)},   \\
\nonumber G_\downarrow = \frac{e^2}{h} |\frac{1}{\hat{\lambda}_\downarrow}|^2 \left( \frac{T}{T_F}
\right)^{2(\frac{1}{{K}_\downarrow}-1)}.
\end{eqnarray}
At low temperature, i.e. $T_F/T \sim 100 - 1000$, we can find that one of the conductances is strongly suppressed which gives rise to a situation that only one spin polarization is conducting.  Consequently, a polarized wire is
obtained.

\section{Numerical Results}
Based on the exact result from equation (21) and (22), we have
plot our resultant ${K}_\uparrow$ and ${K}_\downarrow$ as a
function of the external magnetic field $\Delta$ in figures 1 and
2. In both figures, we have set $g_4 = 0$.  In order to
demonstrate that only the $g_0$ term is important for causing
$K_\uparrow \neq K_\downarrow$, we set $g_2^\perp = 0$ in figure 1
and $g_2^\perp = 1$ in figure 2.

One can understand these two figures from locating in each of the them three pairs of curves starting from $\Delta =0$.  In figure 1, the top most pair
of curves are ${K}_\uparrow$ and ${K}_\downarrow$ represented by ``$\times$'' and ``$\circ$'', respectively.  These are the
results of $g_0 = 4 V = 0$ which correspond to the case when there is no
e-e interaction.  With an arbitrary increase of the external
magnetic field, there is no polarization of the electron gas.  In this case, ${K}_\uparrow = {K}_\downarrow = 1$.  Next, we increase $g_0$ (or $4 V$) to $g_0 = 0.5$.  The $K_\uparrow$ and the $K_\downarrow$ curves are represented by ``$+$'' and ``$\square$'', respectively.  One can see that at $\Delta = 0$, the values of $K_\uparrow$ and $K_\downarrow$ are equal and are a little bit smaller than 1.  By increasing the external magnetic field $\Delta$, we can see that $K_\uparrow$ and $K_\downarrow$ splits, thus giving rise to a polarized electron gas.  Furthermore, the third pair of curves correspond to the case when $g_0 = 1$.  The $K_\uparrow$ and the $K_\downarrow$ curves are represented by ``$\ast$'' and ``$\triangle$'', respectively.  One can clearly see that the $K_\uparrow$ and $K_\downarrow$ curves split further which gives rise to a more polarized electron gas.

In order to show that the $g_2^\perp$ term is indeed not playing a
significant role in the polarization effect, we have plot figure 2
by taking $g_2^\perp = 1$.  By simply comparing these two figures,
one can easily see that the polarization effect is not affected by
the inclusion of the $g_2^\perp$ term.  Notice that the first pair
of curves in figure 2 at $K_\uparrow = K_\downarrow = 1$ does not
split at all, even if we take $g_0 = 0$ and $g_2^\perp = 1$.  For this
reason, we can conclude that only in the presence of the $g_0$
term and $\Delta \neq 0$ can we obtain a polarized wire.

Furthermore, in order to show that the difference between $G_\uparrow$ and $G_\downarrow$ increases when an external magnetic field $\Delta$ is applied, we have prepared figure 3 to figure 6 illustrating the dependency of polarization $ P \equiv |\frac{G_\uparrow - G_\downarrow}{G_\uparrow + G_\downarrow}|$ versus the external magnetic field $\Delta$ and the extended Hubbard interaction $g_0 = 4 V$.

In the case of a weak impurity potential, the resultant polarization is shown in figure 3 and figure 4.  In figure 3, a 3D diagram of the polarization $P$ is plot.  The external magnetic field $\Delta$ is taken as the x-axis, while the extended Hubbard interaction $g_0 = 4 V$ is taken as the y-axis.  The range of $\Delta$ is between $0.$ to $1.6$, while the range of $g_0$ is between $0.$ and $1.$, where $g_2^\perp = 0$, $T = T_F / 500$ and $\hat{\lambda}_\sigma = 0.1$ are taken as typical values.  As shown, the larger the external magnetical field $\Delta$ is applied to the wire and the more the extended Hubbard interaction is turned on, the more the wire is polarized.  Note that the external magnetic field $\Delta$ here is taken in the unit of $v_F = 1$.  In figure 4, a contour plot of figure 3 is provided.  The x-axis represents the external magnetic field $\Delta$ and the y-axis represents the extended Hubbard interaction $4 V$.  Starting from the lower left side of the figure, the contour lines represents the polarization of 5 percent, 10 percent, 15 percent, and so on.  According to our result, the electron gas in the weak impurity potential case is significantly polarized only when the the external magnetic field $\Delta$ and the extended Hubbard interaction $4 V$ are large.

On the other hand, the resultant polarization for the case of a weak link is shown in figure 5 and figure 6.  Similarly, we take the external magnetic field $\Delta$ as the x-axis and the extended Hubbard interaction $g_0 = 4 V$ as the y-axis.  The range of $\Delta$ is between $0.$ to $1.6$, while the range of $g_0$ is between $0.$ and $1.$, where $g_2^\perp = 0$ and $T = T_F / 500$.  In figure 5, the 3D diagram clearly shows that the electron gas is significantly polarized when the external magnetic field $\Delta$ or the extended Hubbard interaction $ 4 V$ is greater than zero.  In figure 6, a contour plot of figure 5 is provided.  Starting from the lower left side of the figure, the contour lines represent 5 percent, 10 percent, 15 percent, etc., of polarization, respectively.  By comparing the results of the weak potential case and that of the weak link case, we can conclude that the stronger the impurity potential, the more the electron gas is polarized.

In real systems and large temperatures, inelastic scattering will give rise to spin relaxation.  As a result, our effect will be drastically reduced.  However, in the ballistic regime, where the length of the wire $L$ is comparable to the thermal length $L_T$, $L \approx L_T \approx 10^{-6} m$ ($ T \approx 2 K^\circ$), our ``spin polarizer'' can be realized.

In summary, the possibility for a magnetic wire and the
possibility for a ``spin polarizer'' caused by magnetic field,
interaction and impurity are suggested.  This effect might play a
significant role in spintronics.

\section{Acknowledgement}
This work is supported by the DOE Grant No. DE-FG02-01ER4909.

\begin{figure}
\includegraphics[width=6.5in]{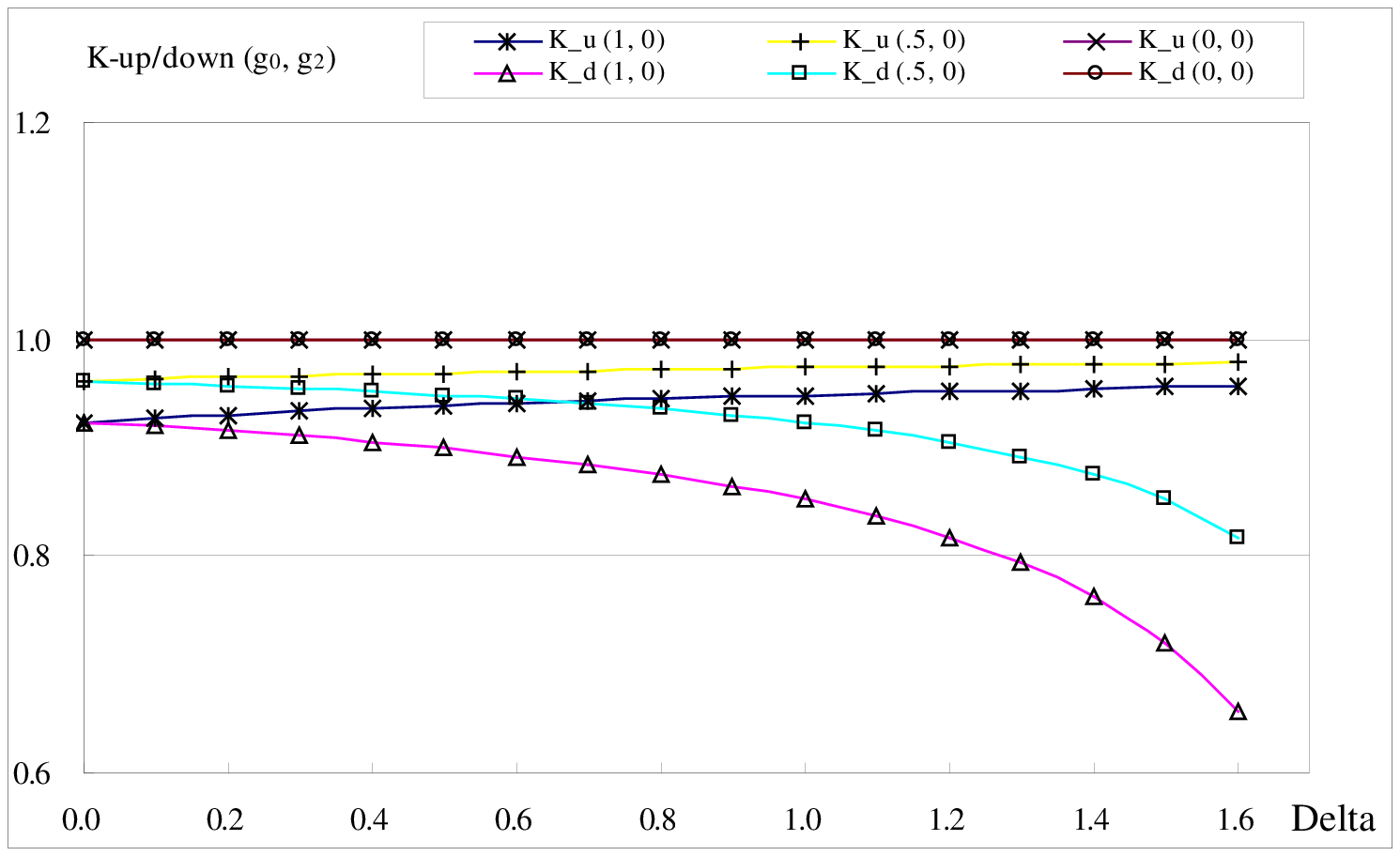}
\caption{The results of $K_\uparrow$ and $K_\downarrow$ by
increasing the external magnetic field $\Delta$ at $g_2^\perp = 0.$
}
\vspace{1cm}
\includegraphics[width=6.5in]{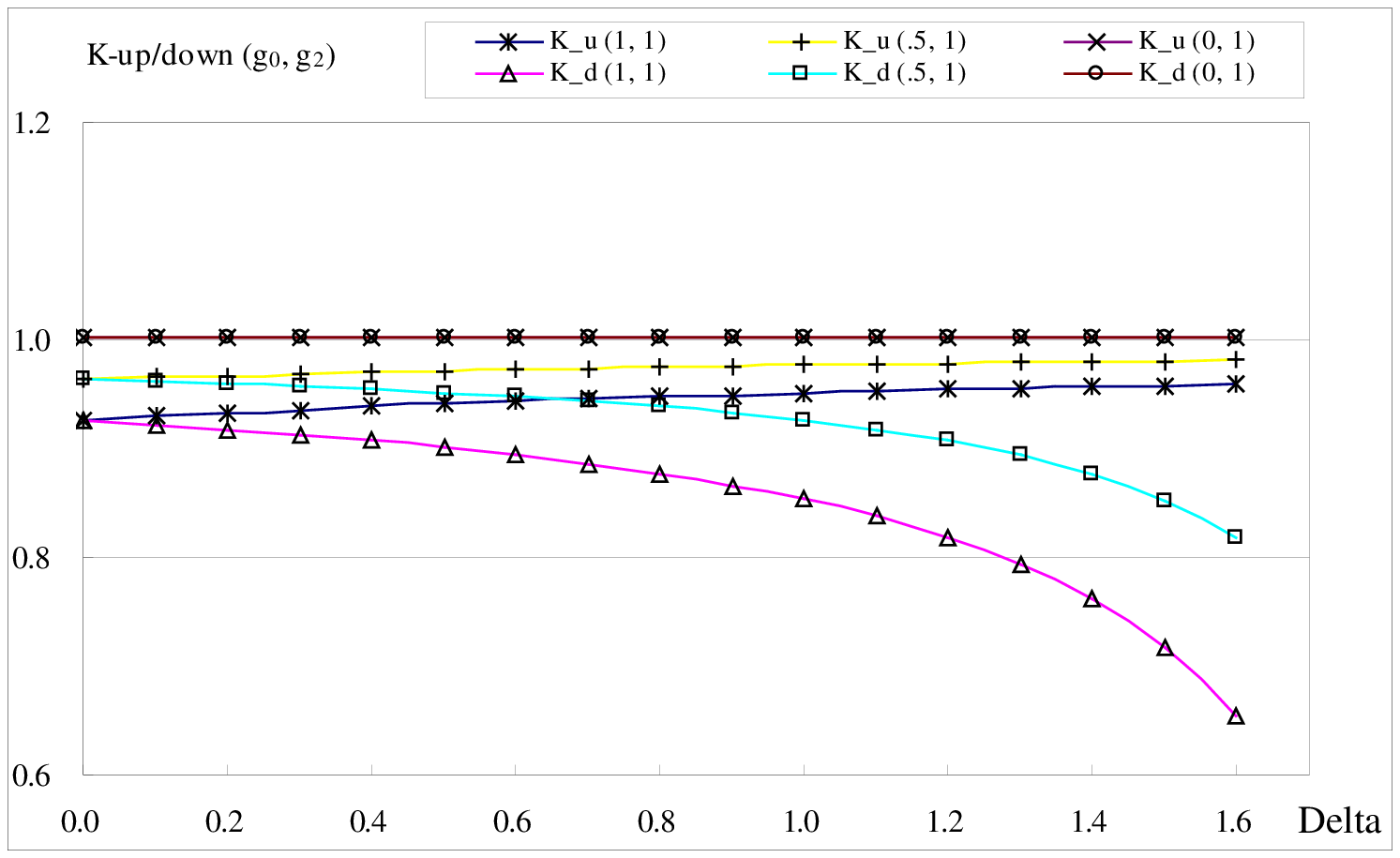}
\caption{The results of $K_\uparrow$ and $K_\downarrow$ by
increasing the external magnetic field $\Delta$ at $g_2^\perp = 1.$
}
\end{figure}
\begin{figure}
\includegraphics[width=3.5in]{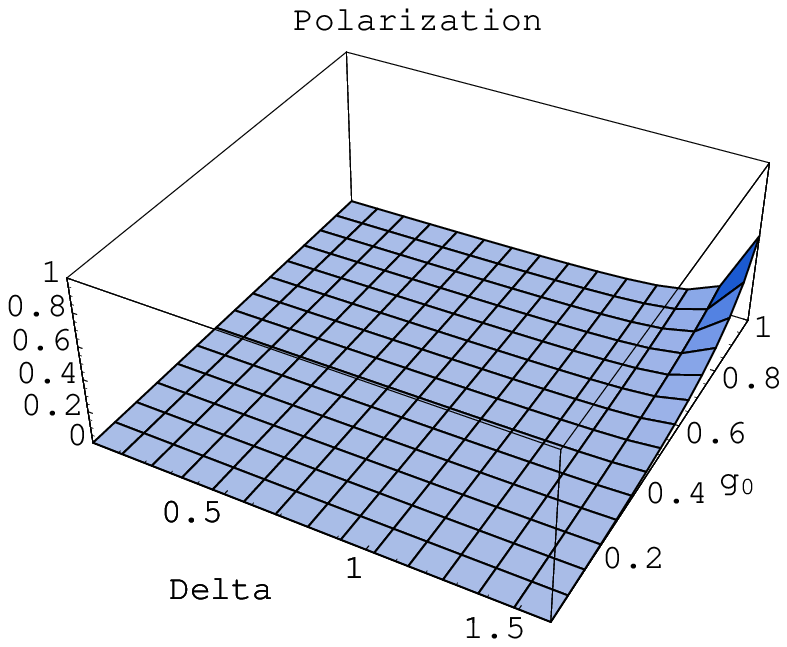}
\caption{This figure shows a 3D diagram of the polarization $P$ in the weak impurity potential limit.  In this figure, we take the external magnetic field $\Delta$ ranging from $0.$ and $1.6$, and the extended Hubbard interaction $g_0 = 4 V$ ranging from $0.$ to $1.$, where $T = T_F / 500$ and $\hat{\lambda}_\sigma = 0.1$ are taken as typical values.
}
\vspace{1cm}
\includegraphics[width=3in]{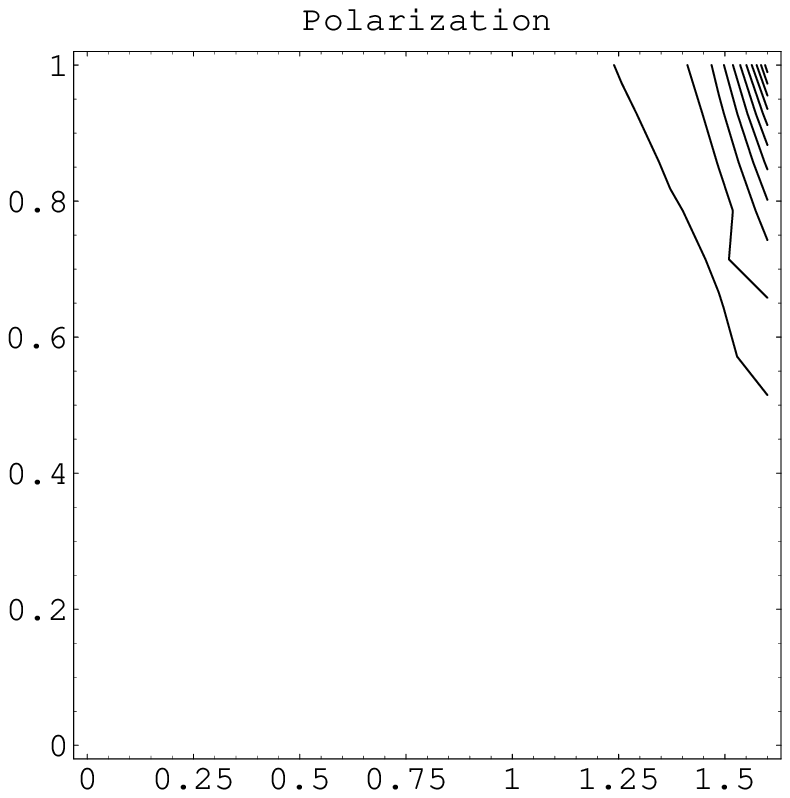}
\caption{This figure shows a contour diagram of polarization $P$ shown in figure 3.  In this figure, the contour line starting from the lower-left side represents the polarization of 5 percent.  The next line represents a polarization of 10 percent, and so on.}
\end{figure}
\begin{figure}
\includegraphics[width=3.5in]{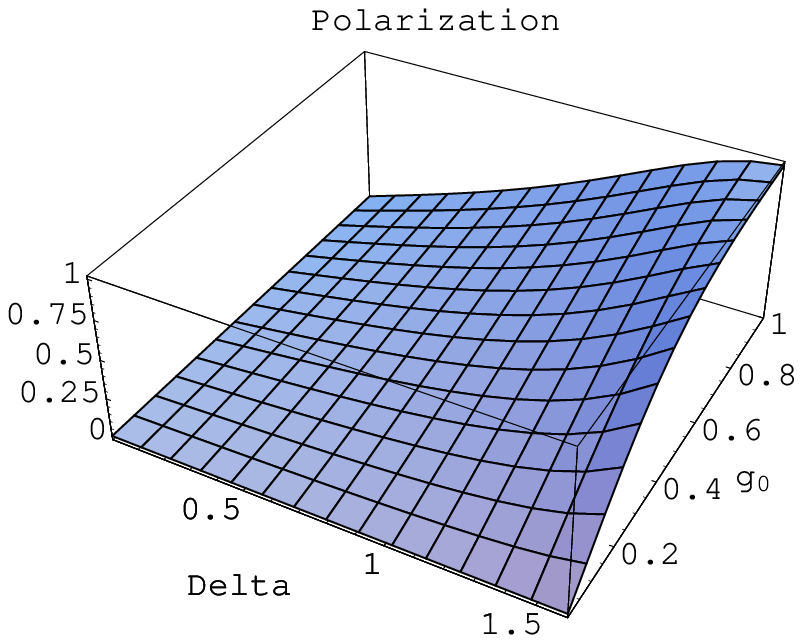}
\caption{This figure shows a 3D diagram of the polarization $P$ in the weak link limit.  In this figure, we take the external magnetic field $\Delta$ ranging from $0.$ and $1.6$, and the extended Hubbard interaction $g_0 = 4 V$ ranging from $0.$ to $1.$, where $T = T_F / 500$ is taken as a typical value.
}
\vspace{1cm}
\includegraphics[width=3in]{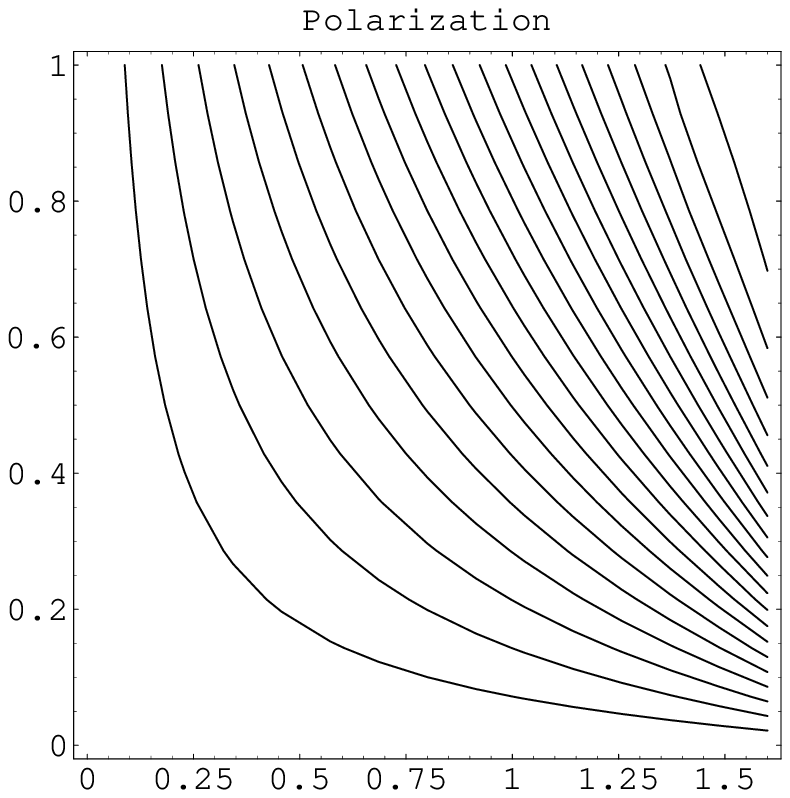}
\caption{This figure shows a contour diagram of polarization $P$ shown in figure 5.  In this figure, the contour line starting from the lower-left side represents the polarization of 5 percent.  The next line represents a polarization of 10 percent, and so on.}
\end{figure}


\begin{thebibliography}{3}
\expandafter\ifx\csname
natexlab\endcsname\relax\def\natexlab#1{#1}\fi
\expandafter\ifx\csname bibnamefont\endcsname\relax
  \def\bibnamefont#1{#1}\fi
\expandafter\ifx\csname bibfnamefont\endcsname\relax
  \def\bibfnamefont#1{#1}\fi
\expandafter\ifx\csname citenamefont\endcsname\relax
  \def\citenamefont#1{#1}\fi
\expandafter\ifx\csname url\endcsname\relax
  \def\url#1{\texttt{#1}}\fi
\expandafter\ifx\csname
urlprefix\endcsname\relax\def\urlprefix{URL }\fi
\providecommand{\bibinfo}[2]{#2}
\providecommand{\eprint}[2][]{\url{#2}}
\bibitem{KaneFisher92PRL}
  \bibinfo{author}{\bibfnamefont{C.L.} \bibnamefont{Kane}},
  \bibinfo{author}{\bibfnamefont{M. P.A.} \bibnamefont{Fisher}},
  \bibinfo{journal}{Phys. Rev. Lett.} \textbf{\bibinfo{volume}{68}},
  \bibinfo{pages}{1220} (\bibinfo{year}{1991}).

\bibitem{Aronov}
  \bibinfo{author}{\bibfnamefont{A.G.} \bibnamefont{Aronov}},
  \bibinfo{journal}{JETP Lett.} \textbf{\bibinfo{volume}{24}},
  \bibinfo{pages}{32} (\bibinfo{year}{1976});
  \bibinfo{author}{\bibfnamefont{A.} \bibnamefont{Brataas et al.}},
  \bibinfo{journal}{Phys. Rev. B} \textbf{\bibinfo{volume}{59}},
  \bibinfo{pages}{93} (\bibinfo{year}{1999});
  \bibinfo{author}{\bibfnamefont{R.} \bibnamefont{Fitzgerald}},
  \bibinfo{journal}{Physics Today},
  \bibinfo{pages}{April} (\bibinfo{year}{2000}).

\bibitem{Schmeltzer02PRB}
  \bibinfo{author}{\bibfnamefont{D.} \bibnamefont{Schmeltzer}},
  \bibinfo{journal}{Phys. Rev. Lett.} \textbf{\bibinfo{volume}{65}},
  \bibinfo{pages}{193303} (\bibinfo{year}{2002}).

\bibitem{LutherEmery}
  \bibinfo{author}{\bibfnamefont{A.} \bibnamefont{Luther}}, and
  \bibinfo{author}{\bibfnamefont{V.J.} \bibnamefont{Emery}},
  \bibinfo{journal}{Phys. Rev. Lett.} \textbf{\bibinfo{volume}{33}},
  \bibinfo{pages}{589} (\bibinfo{year}{1974}).

\bibitem{gogolin}
  \bibinfo{author}{\bibfnamefont{A.}~\bibnamefont{Gogolin}},
  \bibinfo{author}{\bibfnamefont{A.}~\bibnamefont{Nersesyan}}, \bibnamefont{and}
  \bibinfo{author}{\bibfnamefont{A.}~\bibnamefont{Tsvelik}},
  \bibinfo{book}{``Bosonization and Strongly Correlated Systems''},
  \bibinfo{publisher}{Cambridge University Press} (\bibinfo{year}{1998}).

\bibitem{GiamarchiSchulz88PRB}
  \bibinfo{author}{\bibfnamefont{T.} \bibnamefont{Giamarchi}}, and
  \bibinfo{author}{\bibfnamefont{H.J.} \bibnamefont{Schulz}},
  \bibinfo{journal}{Phys. Rev. B} \textbf{\bibinfo{volume}{37}},
  \bibinfo{pages}{325} (\bibinfo{year}{1988}).
\bibitem{KimuraKurokiAoki96PRB}
  \bibinfo{author}{\bibfnamefont{Takashi} \bibnamefont{Kimura}}, 
  \bibinfo{author}{\bibfnamefont{Kazuhiko} \bibnamefont{Kuroki}}, and
  \bibinfo{author}{\bibfnamefont{Hideo} \bibnamefont{Aoki}}, 
  \bibinfo{journal}{Phys. Rev. B} \textbf{\bibinfo{volume}{53}},
  \bibinfo{pages}{9572} (\bibinfo{year}{1996}).
  
\bibitem{193303}
  \bibinfo{author}{\bibfnamefont{David} \bibnamefont{Schmeltzer}},
  \bibinfo{journal}{Phys. Rev. B} \textbf{\bibinfo{volume}{65}},
  \bibinfo{pages}{193303} (\bibinfo{year}{2002}).
  
\end{thebibliography}
\end{document}